# Putting Density Functional Theory to the Test in Machine-Learning-Accelerated Materials Discovery

Chenru Duan[1,2], Fang Liu[1], Aditya Nandy[1,2], and Heather J. Kulik[1,*]

[1]Department of Chemical Engineering, Massachusetts Institute of Technology, Cambridge, MA 02139
[2]Department of Chemistry, Massachusetts Institute of Technology, Cambridge, MA 02139

AUTHOR INFORMATION

**Corresponding Author**

*email: hjkulik@mit.edu, phone: 617-253-4584

ABSTRACT: Accelerated discovery with machine learning (ML) has begun to provide the advances in efficiency needed to overcome the combinatorial challenge of computational materials design. Nevertheless, ML-accelerated discovery both inherits the biases of training data derived from density functional theory (DFT) and leads to many attempted calculations that are doomed to fail. Many compelling functional materials and catalytic processes involve strained chemical bonds, open shell radicals and diradicals, or metal–organic bonds to open-shell transition-metal centers. Although promising targets, these materials present unique challenges for electronic structure methods and combinatorial challenges for their discovery. In this Perspective, we describe the advances needed in accuracy, efficiency, and approach beyond what is typical in conventional DFT-based ML workflows. These challenges have begun to be addressed through ML models trained to predict the results of multiple methods or the differences between them, enabling quantitative sensitivity analysis. For DFT to be trusted for a given data point in a high-throughput screen, it must pass a series of tests. ML models that predict the likelihood of calculation success and detect the presence of strong correlation will enable rapid diagnoses and adaptation strategies. These "decision engines" represent the first steps toward autonomous workflows that avoid the need for expert determination of the robustness of DFT-based materials discoveries.

**TOC GRAPHICS**

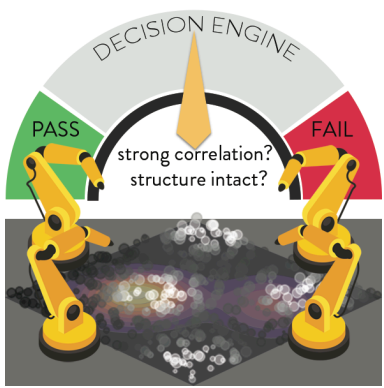

Machine learning (ML) is fast emerging as an essential complement to traditional theoretical chemistry in the accelerated discovery of molecules and materials.[1-3] ML models (e.g., artificial neural networks, ANNs, or kernel ridge regression, KRR) trained on appropriate representations[4] reproduce the results calculated with electronic structure theory to within a suitable margin[5] (ca. 1–3 kcal/mol) in challenging materials spaces such as open-shell transition-metal chemistry[4-7] where lower-fidelity semi-empirical models fail[8-10]. Developments in uncertainty quantification[11-14] ensure application of these ML models only where predictions are well supported by training data. Active learning (i.e., improving ML models on the fly) has accelerated discovery[15-18] by focusing data acquisition to the most fruitful regions of chemical space. These efforts have produced design rules and lead compounds in weeks instead of the decades that conventional high-throughput screening with density functional theory (DFT) would have required[15]. This scale of acceleration is most promising and most needed where traditional Edisonian efforts and intuition have failed and where search is hampered by the vastness of the combinatorial space. Transition-metal complexes exemplify this, as the range of ligands and metals along with variable oxidation and spin state of the metal create a large space for exploration, where quantum mechanical properties are difficult to predict *a priori*. These same challenges for exploration nevertheless represent opportunities or "knobs" for the design of functional materials (e.g., as switches or sensors) and selective catalysts[19-20].

Although the potential payoff for ML-accelerated discovery in challenging chemical spaces is evident, numerous obstacles remain (Figure 1). Low-cost electronic structure methods such as DFT are widely used in the generation of training data for ML models[2-3] and in high-throughput screening[21]. By definition, ML models trained on such data sets will learn the biases of the underlying DFT functional and as a result will magnify these errors in large-scale chemical discovery. For example, accelerated discovery of spin-crossover complexes with a model trained on semi-local DFT will only discover complexes with weaker-field ligands, whereas one trained with hybrid DFT will identify those with stronger-field ligands.[22] While in small-scale studies the validity of each DFT calculation might be inspected by hand, within large-scale automated workflows[21,23-25] manual inspection of calculation outcomes (e.g., to determine if a structure remains intact) is not possible. In addition, expert knowledge is required to determine how to accurately model systems with strong correlation[26-27] or to tell when the wavefunction has converged to an unexpected (e.g., broken symmetry) result. For computational materials design[28] with ML[1-3] to deliver on its promise, we must bring expert computational chemistry knowledge into the accelerated discovery loop at the speed that ML models can deliver. In this Perspective, we describe progress towards the goal of building fully autonomous computational chemistry workflows (Figure 1). The artificial intelligence "decision engines" discussed here complement and augment expert knowledge of the computational chemist, enabling drastic increases in the size of chemical space that can be explored.

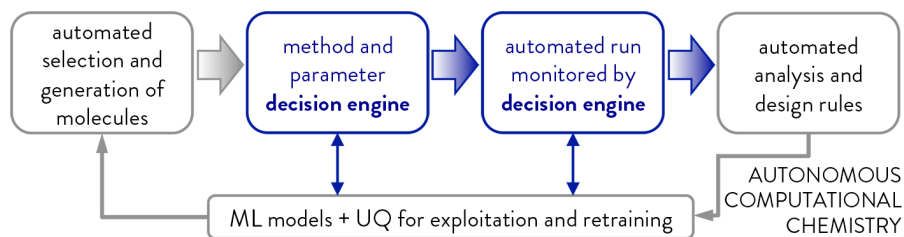

**Figure 1.** Schematic of the steps required in a fully autonomous computational workflow for chemical and materials discovery. The steps with widely available, established tools are shown in gray: automated selection and generation, analysis, and development of machine learning (ML) models with uncertainty quantification (UQ). The missing links for autonomous computational chemistry are shown in dark blue: artificial intelligence decision engines to select electronic structure methods and to enable monitoring of successful calculations.

A chief impediment for ML-accelerated screening of challenging chemical spaces is the high rate of failure of attempted calculations.[7,29] For open-shell transition-metal complexes, we classify as "failure" any of the following: changes in metal coordination number or isomer, rearrangement of atoms within or between ligands, and convergence of the wavefunction to an unexpected electronic state or with a large degree of spin contamination (Figure 2).[7] Our heuristic definition for significant spin contamination is when deviations of the $<S^2>$ operator from its expected $S(S+1)$ value by greater than 1 $\mu_B^2$, motivated by observations in Ref. 30 that deviations of this size led to low spin states that effectively had the properties of higher spin states in small transition metal complexes. While a computational chemist can inspect changes in structure manually and adjust the study appropriately, machine learning techniques using graph-based representations[4] require an unambiguous chemical structure.[7] Spin contamination and the lack of contribution from the metal center to the overall spin are also signs to a computational chemist that higher levels of theory might be needed or that a ligand is redox noninnocent[31-32]. ML models trained to predict spin-state-dependent properties from DFT require a more conservative filter in the data generation step to ensure the same qualitative states are being compared across the dataset.

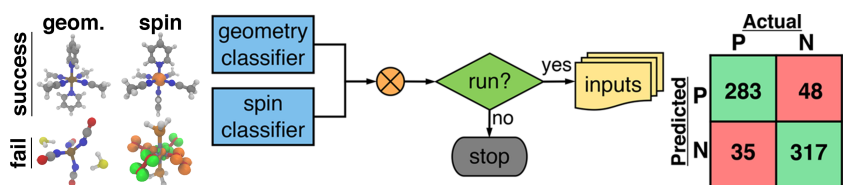

**Figure 2.** (Left) Examples of successful and failed calculations according to spin and geometric (geom.) classifications. (Middle) Workflow for combined spin and geometric static classifier for aborting calculations. (Right) Confusion matrix of positive (i.e., successful) and negative (i.e., failed) combined spin and geometry outcomes with a 0.5 threshold from each ANN classifier. Adapted from Ref. 29. Copyright 2019 American Chemical Society.

Compounding the challenge of calculation failure is that most optimization targets involve multiple calculations, e.g., several intermediates in a catalytic cycle as well as variable spin/oxidation state or total charge. Thus, failure of any mode of one calculation results in total loss of property evaluation. Whether in one-shot model training[4,7,33-34] or active learning[15-17], the result is that less data is acquired. In a recent example[15], redox couple prediction, which requires three successful geometry optimizations, saw success rates drop from 35% to 20% after five generations of active learning despite a high (50–65%) individual calculation success rate.[15]

Some calculation failures are foreseeable, but others are less so. A relatively free, automated combinatorial exploration of chemical space could be anticipated to produce incompatible combinations of ligands that are too highly charged, bulky, or prone to non-innocence. Still, many failures are less intuitive and may, for a given complex, depend on spin or oxidation state.[29] This mapping between chemistry and calculation failure is amenable to machine learning, and we trained[29] ANN classifiers to accurately (ca. 88–95%) predict the likelihood of calculation failures using graph-based revised autocorrelation (RAC) functions[4,35-36] (Figure 2 and Table 1). Training a classifier to predict calculation failure based on chemical composition has the advantage that the trained model can predict success in less than a second, avoiding costly DFT calculations that are likely to fail (Figure 2). Since we observed that calculations take longer to fail than to succeed[29], one both avoids generating spurious data for property prediction models and most (ca. 88%) of the feasible chemical space is explored in 33% of the computational time (Table 1).[29]

**Table 1.** The accuracy and area under the curve (AUC) of ANN models for geometry and spin classification tasks, as judged on the validation set used for hyperparameter selection, the set-aside test set, and an out-of-distribution set consisting of metal–oxo catalytic complexes. Adapted from Ref. 29. Copyright 2019 American Chemical Society.

| Metric | Validation | Test | Oxo set |
|---|---|---|---|
| **Geometry** | | | |
| Accuracy | 0.88 | 0.88 | 0.75 |
| AUC | 0.95 | 0.95 | 0.66 |
| **$<S^2>$ deviation** | | | |
| Accuracy | 0.95 | 0.96 | -- |
| AUC | 0.98 | 0.97 | -- |

Despite their promise, composition-derived (i.e., RAC-based) classifiers have some limitations. Failed calculations must be intentionally generated or retained. It is wise to preserve such meta-data in high-throughput screening, as the need to anticipate how failed results bias discovery is still being understood throughout chemistry (e.g., in drug[37] and materials discovery[38-39]). However, the overhead of up-front data generation is overshadowed by the concern of transferability. Changing oxidation state and chemistry, e.g., from complexes derived from the spectrochemical series to metal-oxo species relevant for catalysis, erodes classifier performance to little better than random guess (Table 1). Uncertainty quantification based on the point's perceived (i.e., in the model latent space) distance to training data[11,29] can be used to limit predictions to regions where the model is confident, but this will lead to predictions for a smaller fraction of the space.

We took an alternative approach, similar to recent efforts in transferable machine learning for property prediction[40-42], by noting that the electronic and geometric structures evolve during the geometry optimization. As a calculation proceeds, metrics of the geometric health of the complex[7] (e.g., distances of ligands to the metal) or spin contamination indicate failure of a calculation, and failure or evidence of an impending failure becomes more evident with increasing numbers of steps in the optimization. A set of around 30 electronic (i.e., bond order, gradient, and partial charge) and geometric[7] descriptors from the initial steps (ca. 2–40) of geometry optimizations were used to train convolutional neural networks (CNNs) to predict calculation failure (Figure 3).[29] Even for the model given the most information about optimizations, 40 steps represents only about 6 hours of GPU-accelerated calculation time[43] and still no more than one third of the total average time of a typical calculation.[29] As a demonstration of its power, we recently incorporated an extension of the dynamic classifier introduced in Ref. 29 into workflows for active learning[15] of methane-to-methanol catalysts[44] in order to reduce calculation attrition rates in catalyst discovery (Texts S1–S2, Table S1, and Figures S1–S5). The extended "dynamic" model is a multi-task network that predicts three measures of calculation health (i.e., metal spin, deviations of $<S^2>$ from expected values, and geometry optimization outcome) based solely on properties of the wavefunction and the energetic gradient, without explicit information about chemical composition (Figure 3). As a result, this model represents a more transferable approach[40-42] that generalizes better to new, difficult, and out-of-distribution transition-metal complexes than the "static", compositional approach did.[29] From limited training data, these models efficiently avoid at least half of all failed calculations (Texts S1–S2 and Figures S2 and S5).

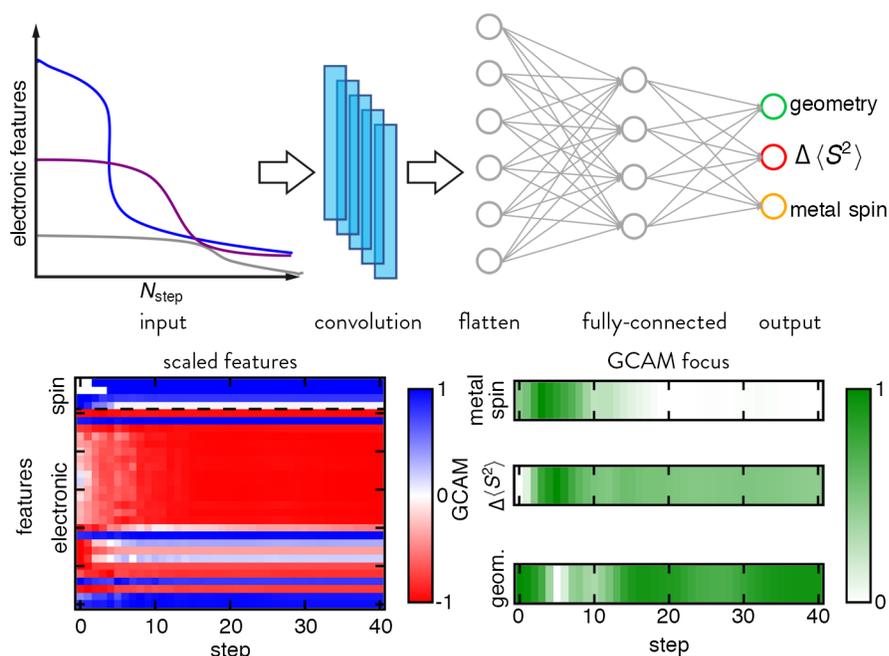

**Figure 3.** (top) Schematic of how electronic features are collected from the properties of the wavefunction during a geometry optimization and used as input to an ANN classifier with a convolutional layer followed by a fully connected layer. The model is a multi-task ANN classifier that predicts calculation success with respect to geometry, deviations of $<S^2>$ from the expected value, and the deviation of the spin on the metal from the total spin. (bottom) Gradient-weighted class-activation map (GCAM)[45] analysis of the multi-task model. In this example, the geometry optimization leads to a good result for all three properties, but features oscillate in early stages of the optimization (left). For each of the three prediction models, the GCAM focus is on different phases of the geometry optimization (right).

When paired with uncertainty quantification metrics (i.e., measured based on the latent space[29]), the model achieves high (ca. 95–99%) accuracy by making predictions only on the complexes for which it is most confident (Figure S3). The model confidence increases with information from a greater number of optimization steps (Figure S3). CNNs can be interrogated to identify which aspect of the features have the greatest influence on predictions, for example by using a gradient-weighted class-activation map[45]. With this analysis, it becomes clear why the "dynamic" approach outperforms the "static" one (Figure 3 and Figure S4). The optimizations exhibit path dependence, with some taking more or less time to fail or succeed in a manner that would be difficult to predict based on composition alone. In the future, models that use the information generated during calculations will be used to adjust calculation parameters on the fly to reduce failure rates further with dynamic control.

In ML-accelerated high-throughput screening of challenging targets (e.g., in transition-metal chemistry) with approximate electronic structure methods, it is seldom known *a priori* how the discovery process will be influenced by the choice of method. Here, approximate DFT is applied nearly exclusively, and DFT functionals that perform well for one class of materials[46] often fail to predict properties of another.[47] While it is reasonable to tailor a functional to predict properties across a narrow range of chemical space (e.g., Fe(II)/N spin-crossover complexes[46]), broader discovery applications will not have experimental benchmark data or high-accuracy electronic structure results against which the optimal functional can be selected. Similarly, when traversing large regions of chemical space, one can anticipate that some regions of the space are more sensitive to method choice than others.[22] Open-shell transition-metal chemistry is an example of a method-sensitive region of chemical space, where any "Jacob's ladder" hierarchy fails[48-51] due to delicate trade-offs between static correlation error[52] that is larger in hybrid functionals[52-53] and delocalization error evident in pure functionals[52-54].

Pragmatic, low-cost approaches to addressing shortcomings of DFT or single-reference wavefunction methods are essential for high-throughput screening. Techniques relying on cancellation of errors, e.g., by adjusting between experiment and computational predictions for a reference compound[55-56], cannot necessarily be guaranteed to work across the wide range of chemical compounds studied in a high-throughput screen. While not yet demonstrated for practical high-throughput screening, game theory has been proposed as a means to select[57] optimal functional/basis set pairs[58]. In game theory, strategies are selected that maximize the overall benefit to multiple players. For DFT functional and basis set selection, knowledge of functional/basis set performance across a benchmark set as well as chemical similarity of a new molecule to these benchmark data was proposed as a three player problem (i.e., method accuracy, calculation complexity/cost, and chemical similarity) in game theory.[58] Sensitivity analysis and uncertainty quantification (UQ) are two complementary approaches that have been fruitfully applied in transition-metal catalysis, where the optimal electronic structure method is often unknown. Bayesian inference has been used to curate a series of DFT functionals for use in surface science[59-60] by creating an ensemble of exchange-correlation functionals[59] or by using a selection of popular off-the-shelf functionals[60]. For open-shell transition-metal chemistry, the fraction of Hartree–Fock (HF) exchange plays a critical role in property (e.g., spin state ordering[61-66]) prediction. Since properties vary strongly but linearly with HF exchange, HF exchange sensitivity analysis[67] provides a useful guide to which regions of chemical space are most sensitive to functional choice[22]. Using ML models that predict properties at variable HF exchange fractions or that directly predict the sensitivity, we have visualized how lead compounds in large chemical spaces (e.g., over 5k spin-crossover complexes[22,68]) change with changing functional (Figure 4). These ML model predictions encapsulate expectations (e.g., that GGAs have a low-spin bias that favors weaker-field ligands, whereas hybrids reverse this tendency) and also highlight regions of space with properties that are relatively insensitive to functional choice (Figure 4). Given the ability of ML models to be trained on and to predict relationships between distinct functionals, one can imagine carrying out chemical discovery objectives while requiring consensus amongst functionals or methods. Although consensus does not guarantee that the electronic structure is correct, it does reduce a seldom-addressed source of uncertainty.

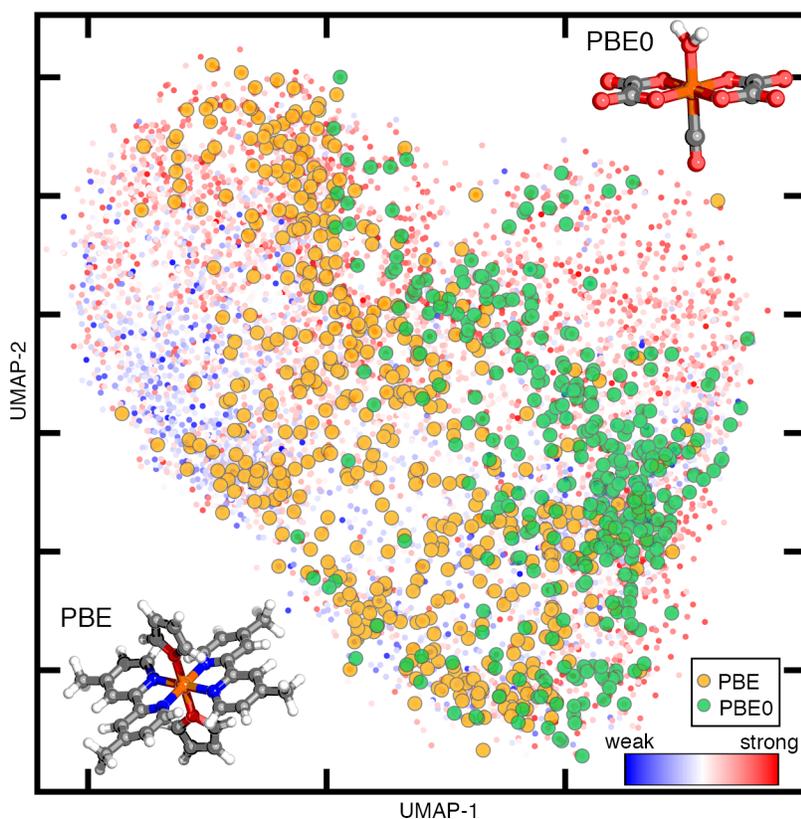

**Figure 4.** (top) A uniform manifold approximation and projection (UMAP)[69] visualization of a 10% random sample of 187k hypothetical complexes from Ref. [70]. Each complex (dots) is colored by a heuristic measure of ligand field strength from the average coordinating atom electronegativity (blue to red, see inset colorbar). Spin-crossover (SCO) complexes discovered by ANNs trained to predict pure GGA (PBE, in orange) or hybrid GGA (PBE0, in green) spin-splitting (i.e., $\Delta E_{H-L}$) energies are visualized on top of this space. As in Ref. 22, an SCO is defined as $|\Delta E_{H-L}| < 2.5$ kcal/mol. Example SCOs for each functional are shown in insets:

intermediate-field Fe(III)(4,4'-dimethylbipyridine)$_4$(furan)$_2$ for PBE and a weaker-field Fe(III)(oxalate)$_2$(H$_2$O)(CO) for PBE0.

One way to sidestep the choice of DFT functional is to use higher-scaling correlated wave function theory (WFT) methods. Single-reference coupled cluster methods are the gold standard in organic chemistry. However, coupled cluster can be expected[27,71] to perform as poorly as DFT or worse in cases where strong correlation is present. In these cases, it may be necessary to carry out time-consuming, tedious multi-reference (MR) calculations. Although some systematic approaches have been proposed[72-74] for active space and orbital selection in MR methods, MR WFT is still too computationally demanding to apply in high-throughput screening. A number of diagnostics have been developed[75-89] to probe for the presence of MR character, but they tend to disagree except in extreme cases.[90] Thus, practitioners often choose a combination of diagnostics[26,78,91], adjust the thresholds for diagnosing strong correlation based on material[26], or otherwise rely on experience to make these decisions. These less systematic behaviors can be challenging to replicate in an automated workflow.

Complicating their use in high-throughput workflows is the fact that the MR diagnostics that are highly accurate (e.g., the leading weight of the CASSCF wavefunction[76,85-87] or amplitudes from CCSD[76-78]) are both high-cost and require nearly as much care as the full MR WFT calculation, making them tractable, e.g., in transition-metal chemistry, for only the smallest molecules[26,92]. Fractional occupation number(FON)-based diagnostics from DFT provide intuitive (i.e., through spatial representations of the fractional occupation density) qualitative interpretations of MR character at low cost.[79-80,89] Much like the HOMO–LUMO gap of a transition-metal complex[7,93], quantitative measures derived from finite-temperature DFT (i.e., FON-based diagnostics) can be learned by ANNs[70] trained on graph-based RAC descriptors[4].

Interestingly, despite both quantities being derived from the placement of frontier orbitals, FON-based measures of MR character do not have a good linear correlation to HOMO–LUMO gaps (Figure 5). An advantage of training ML models to predict both quantities in chemical discovery is that we could train on a modest number of available data points, e.g., ca. 4,000 complexes from prior studies[6-7,15,29,68], stored in a central repository. We then predicted and visualized a much larger space (here, 187k) of hypothetical compounds to identify "DFT safe" islands, even in cases (e.g., small HOMO–LUMO gap) where one would expect DFT to fail. These methods can also be used to validate or refute expectations of imbalances in strong correlation between spin states[94-95], which are more apparent for some metals (e.g., Cr(II)) than others (e.g., Fe(II)) (Figure 5).[70]

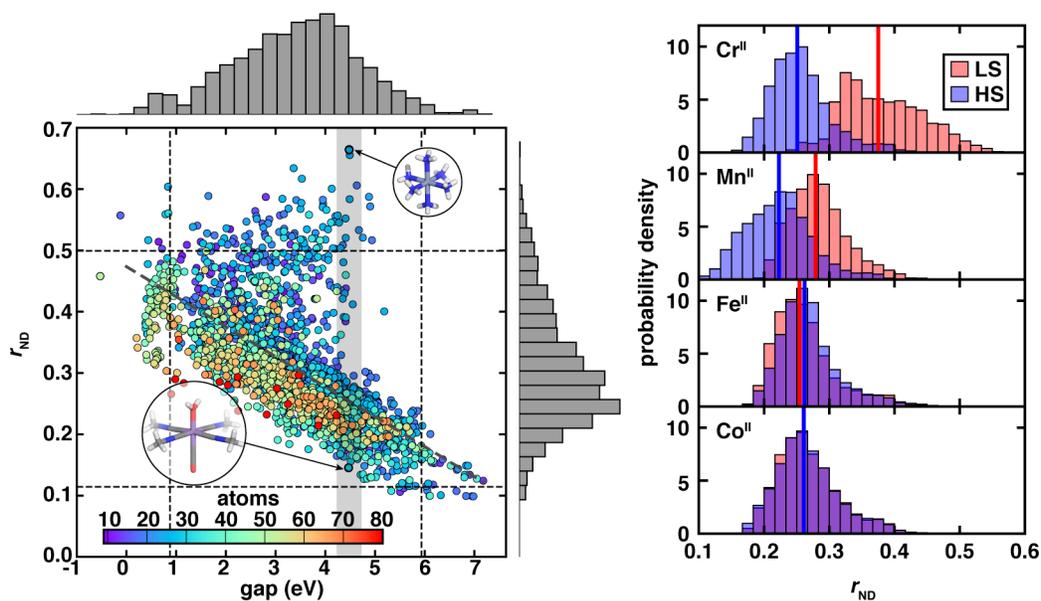

**Figure 5.** (left) HOMO–LUMO gap (in eV) vs $r_{ND}$ (i.e., the ratio of non-dynamical to total correlation from FON DFT) values ($R^2 = 0.41$) for a set of over 2600 transition-metal complexes curated in Ref. [70]. Symbols are colored by number of atoms (inset colorbar), 1D histograms are shown, and dashed lines correspond to two standard deviations around the set mean. For the gray shaded region (gap around 4.5 eV), complexes with the highest (LS $Cr^{II}(NH_3)_6$) and lowest (HS $Mn^{II}(misc)_4(H_2O)(CO)$) $r_{ND}$ values are shown. (right) Normalized probability density distribution of $r_{ND}$ (median shown as a vertical line) predicted by an ANN trained in Ref. [70] on 93.6k theoretical M(II) complexes (M=Cr, Mn, Fe, Co). The translucent histograms are colored red for low-spin and blue for high-spin. Adapted from Ref. [70]. Copyright 2020 American Chemical Society.

Despite the promise of ML models to directly predict these lower-cost DFT-based diagnostics to accelerate high-throughput screening, they are not a perfect solution. The DFT-based MR diagnostics have a tendency to be less predictive of MR character than higher-cost WFT-based diagnostics (Figure 6). However, poor linear correlations between DFT-based diagnostics and the target description of strong correlation can be overcome by using ML models with lower-cost diagnostics as inputs to predict high-cost (i.e., WFT-based) diagnostics[90], similar to transfer learning[41-42,96]. Additional motivation for such an approach comes from the observation that all diagnostics agree about the most extreme SR or MR molecules but tend to disagree about intermediate cases.[70,90] To augment the DFT-based diagnostics, we used size-independent, structure-dependent (i.e., CD-RAC[90]) chemical descriptors to encode the bonding environments that are most strongly associated with MR character. Indeed, ML (i.e., KRR) models trained on a combination of DFT diagnostics and CD-RAC descriptors accurately predict the WFT-based diagnostics at a DFT-level cost amenable to high-throughput screening (Figure 6).

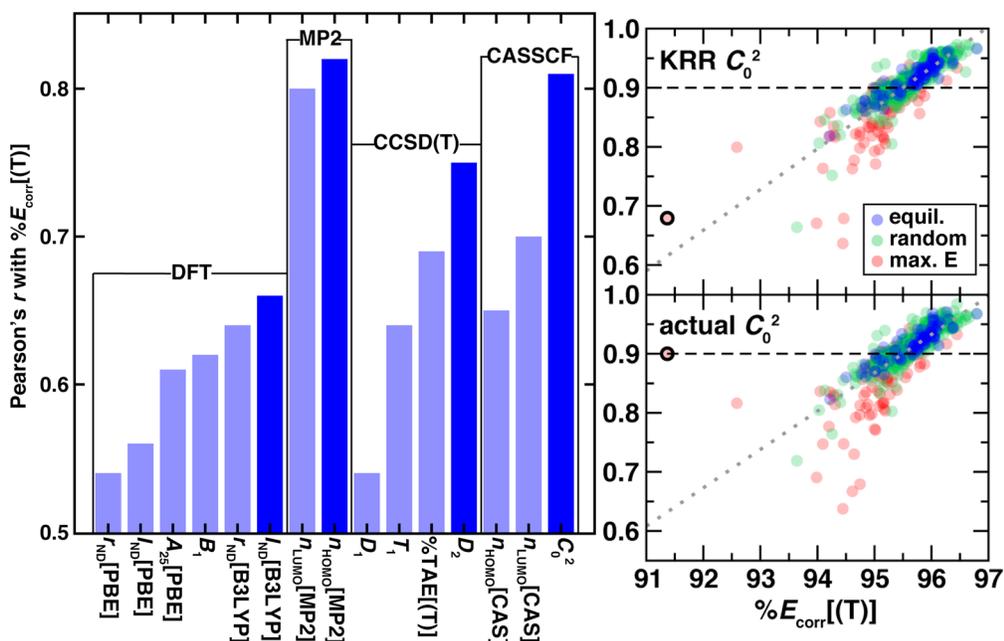

**Figure 6.** (left) Bar graph of unsigned Pearson's $r$ values for percent recovery of the CCSD(T) correlation energy by CCSD, $\%E_{corr}[(T)]$, with 15 MR diagnostics over a set of 3,165 organic structures grouped by level of theory. The best-performing diagnostic is shown opaque and the remainder translucent. (right) $\%E_{corr}[(T)]$ vs actual $C_0^2$ (bottom) or KRR-predicted $C_0^2$ (top). Points are classified as equilibrium (blue circles), randomly distorted (green circles), or highest-energy (max. E, red circles) structures, and the linear correlation (dotted gray line) is computed over the complete set. A suggested MR cutoff (< 0.9) is shown as a black dashed line, and the largest KRR model error point is shown with a black outline. Adapted from Ref. [90]. Copyright 2020 American Chemical Society.

While ML models can be trained to predict higher-accuracy WFT-based diagnostics, the WFT-based diagnostics themselves still often disagree about MR character of individual molecules. For instance, cutoff-based assessments of strong correlation using MP2 occupation numbers would lead one to conclude over a set of 3,165 equilibrium and highly distorted molecules that almost none are MR, whereas excitation-based diagnostics would classify a larger fraction as MR (Figure 7). This disagreement occurs despite both diagnostics providing a good linear correlation with estimates of MR character (i.e., % recovery of the correlation energy from CCSD vs CCSD(T) or CCSDT, Figure 7).[70,90] The picture of consensus at extreme points motivated us to use a semi-supervised machine learning approach known as virtual adversarial training (VAT). In this approach, one trains an ANN with a modified loss function that contains both a supervised and an unsupervised term. Only the most extreme points are labeled, and the classifications of other points are learned during model training (Figure 7). VAT models are robust to noisy inputs, meaning that ML-model-predicted MR diagnostics with a small amount of noise are suitable inputs to the VAT model (Figure 7). In combination, this produces a robust "decision engine"[70] that identifies when MR character is present in a manner that can distinguish molecules where MR methods are required from those where SR methods are sufficient (Figure 7). Currently, this approach has only been trained on closed-shell CNOPS-containing small molecules, and the next step is to extend it to transition-metal complexes. Such extensions may need to exploit real-space diagnostics[79-80,89] in conjunction with bond additivity[97] or composite[98] methods to identify and account for MR character hot spots (i.e., that could be metal- or ligand-centered).

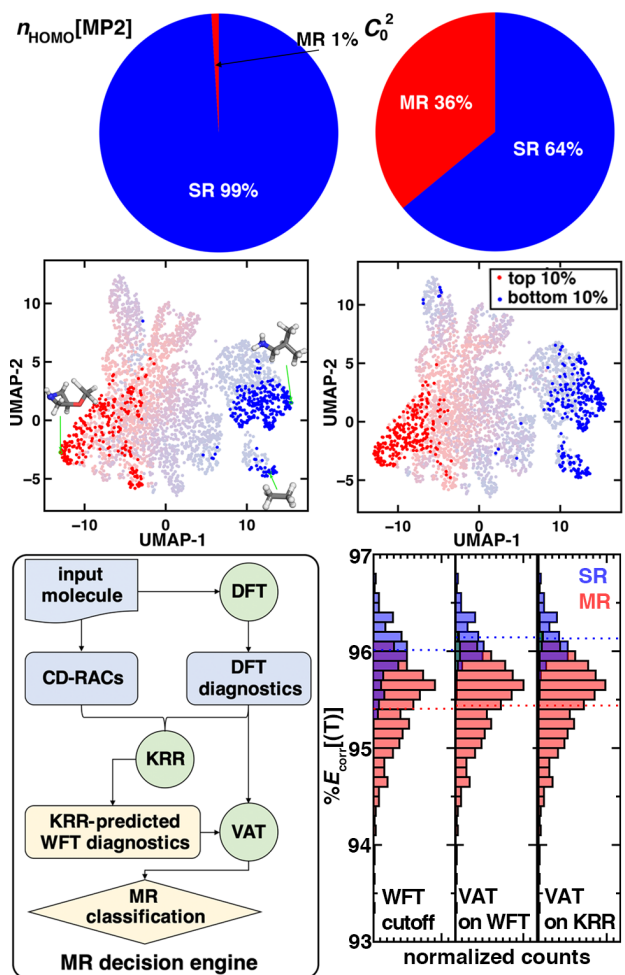

**Figure 7.** (top) MR/SR character of 3,165 organic structures assigned with cutoffs based on $n_{\text{HOMO}}$[MP2] (left) and $C_0^2$ (right). Below each pie chart, a UMAP visualization of all 15 MR diagnostics with the structures in the top 10% (red) and bottom 10% (blue) of the relevant MR diagnostic ($n_{\text{HOMO}}$[MP2] at left and $C_0^2$ at right) are shown as solid circles, with representative structures shown in inset. (bottom, left) Flowchart of the MR decision engine that combines DFT-calculated diagnostics with CD-RACs as input to KRR models to predict WFT MR diagnostics, which are used alongside DFT-calculated diagnostics as input into the VAT model for MR classification. (bottom, right) Overlapping normalized histograms of %$E_{\text{corr}}$[(T)] in SR (blue), transition (green), and MR (red)-classified structures shown for: classification based on a single WFT-based cutoff (left pane), VAT model from WFT and DFT-based diagnostics (middle), and VAT model on KRR-predicted WFT diagnostics (right pane). The SR and MR distributions overlap less for the VAT model than they do for the WFT cutoff model, and the two VAT models behave comparably. Adapted from Ref. [70]. Copyright 2020 American Chemical Society.

In summary, the promise of ML-accelerated materials discovery exemplified by open-shell transition-metal chemistry is matched by the challenge to move beyond the typical accuracy, efficiency, and approach of conventional DFT-based workflows. While ML models have accelerated the discovery process, the effect of bias inherited from the training data source is not well understood. A machine learning approach similar to the one that has accelerated property prediction has been leveraged to predict quantities such as the likelihood of calculation failure or the presence of strong correlation, where chemical intuition or simpler models would fail. Once integrated into high-throughput screening workflows, these models afford improvements needed to accelerate the exploration of computationally challenging chemical spaces. While it is clear that lower level theories are in many cases not adequate for obtaining materials that achieve design objectives, it is less clear if design principles obtained from lower level theories are robust. If the design principles hold, they could be employed to accelerate materials search with higher level theories in the same spirit as transfer learning has demonstrated promise in organic chemistry.[41-42] Rapid, consistent extraction of unbiased experimental reference data[99] will aid assessment of ML predictions. While challenges remain in directly applying MR methods in high-throughput workflows, developments will continue to improve their speed[100-102] and black-box nature[72]. Machine learning will contribute to improving the performance and reducing the cost of high-scaling methods to make higher-level data generation tractable for larger molecules.[40,103-104] Once these pieces come together, fully autonomous computational chemistry will bring predictive ML-accelerated computational discovery to fruition.

ASSOCIATED CONTENT

**Supporting Information**. Extended description of modified dynamic classifier architecture and features, performance, and detailed analysis. (PDF)

This material is available free of charge via the Internet at http://pubs.acs.org.

AUTHOR INFORMATION

**Notes**

The authors declare no competing financial interests.

**Biographies**

**Chenru Duan** is a Ph.D. candidate in Chemistry in the group of Professor Heather J. Kulik at MIT. Prior to that, he obtained his B.S. in Physics from Zhejiang University in China in 2017.

**Fang Liu** is an Assistant Professor in Chemistry at Emory University. Prior to that, she did postdoctoral research in the group of Professor Heather J. Kulik at MIT (2017–2020) and obtained her Ph.D. in Chemistry at Stanford in the group of Todd J. Martínez in 2017.

**Aditya Nandy** is a Ph.D. candidate in Chemistry in the group of Professor Heather J. Kulik at MIT. Prior to that, he obtained his B.S. in Chemical Engineering from UC Berkeley in 2017, with a minor in Chemistry.

**Heather J. Kulik** is an Associate Professor in the Department of Chemical Engineering at MIT. She received her B.E. in Chemical Engineering from the Cooper Union in 2004. She obtained her Ph.D. from the Department of Materials Science and Engineering at MIT in 2009 in the group of Nicola Marzari. She completed postdocs at Lawrence Livermore with Felice Lightstone and Stanford with Todd Martínez, prior to joining MIT as a faculty member in 2013.

ACKNOWLEDGMENT

The authors acknowledge support by the Office of Naval Research under grant numbers N00014-17-1-2956, N00014-18-1-2434, and N00014-20-1-2150, DARPA grant D18AP00039, the Department of Energy under grant numbers DE-SC0018096 and DE-SC0012702, the National Science Foundation under grant numbers CBET-1704266 and CBET-1846426, and MIT Energy Initiative seed grants (2014, 2017). H.J.K. holds a Career Award at the Scientific Interface from the Burroughs Wellcome Fund and a Marion Milligan Mason Award from the AAAS, both of which supported this work. Some of the work discussed in this Perspective made use of Department of Defense HPCMP computing resources or the Extreme Science and Engineering Discovery Environment (XSEDE), which is supported by National Science Foundation grant number ACI-1548562. The authors thank Adam H. Steeves for providing a critical reading of the manuscript.